\documentclass[prl,showpacs,preprintnumbers,preprint,amsmath,amssymb]{revtex4}%
\usepackage{graphicx}
\begin{document}
\bibliographystyle{apsrev}
\title{Evidence for large electric polarization from collinear magnetism in TmMnO$_3$}

\author{V.~Yu. Pomjakushin$^{1}$, M. Kenzelmann$^{1,2}$, A.
D\"{o}nni$^{3}$, A.~B. Harris$^{4}$, T. Nakajima$^{5}$, S.
Mitsuda$^{5}$, M. Tachibana$^{6}$, L. Keller$^{1}$, J. Mesot$^{1}$,
H. Kitazawa$^{3}$, E. Takayama-Muromachi$^{6}$}

\affiliation{(1) Laboratory for Neutron Scattering, ETH Z\"{u}rich
\& Paul Scherrer Institute, CH-5232 Villigen, Switzerland\\(2)
Laboratory for Solid State Physics, ETH Z\"{u}rich, CH-8093
Z\"{u}rich, Switzerland\\(3) National Institute for Materials
Science (NIMS), 1-2-1 Sengen, Tsukuba, Ibaraki 305-0047,
Japan\\(4)Department of Physics and Astronomy, University of
Pennsylvania, Philadelphia, Pennsylvania 19104, USA\\(5) Department of Physics, Faculty of Science, Tokyo University of Science, 1-3 Kagurazaka,
Shinjuku-ku, Tokyo 162-8601, Japan\\(6)	National Institute for Materials Science (NIMS), 1-1 Namiki, Tsukuba, Ibaraki 305-0044, Japan}

\date{\today}

\begin{abstract}
There has been tremendous research activity in the field of
magneto-electric (ME) multiferroics after Kimura {\it et al.}
\cite{Kimura} showed that antiferromagnetic and ferroelectric order
coexist in orthorhombically distorted perovskite ${\rm TbMnO_3}$ and
are strongly coupled. It is now generally accepted that
ferroelectricity in ${\rm TbMnO_3}$ is induced by
magnetic long range order that breaks the symmetry of the crystal
and creates a polar axis \cite{KenzelmannTbMnO3}. One remaining key
question is whether magnetic order can induce ferroelectric polarization that is as large as that of technologically useful materials. We
show that ferroelectricity in orthorhombic (o) ${\rm TmMnO_3}$ is
induced by collinear magnetic order, and that the lower limit for its
electric polarization is larger than in previously investigated
orthorhombic heavy rare-earth manganites. The temperature
dependence of the lattice constants provides further evidence of large
spin-lattice coupling effects. Our experiments suggest that the
ferroelectric polarization in the orthorhombic perovskites with
commensurate magnetic ground states could pass the $1\mathrm{\mu C/cm^2}$
threshold, as suggested by theory \cite{SergienkoPRL,Picozzi}.
\end{abstract}
 \maketitle

Multiferroic materials are defined as materials with more than one switchable spontaneous order parameter such as ferromagnetism and ferroelectricity. It has become custom to include materials with coexisting spontaneous antiferromagnetic and ferroelectric order in the class of ME multiferroics. One can distinguish two major classes of ME multiferroics: those where the onset of ferroelectricity is unrelated to magnetic order, and those where ferroelectricity is induced by magnetic order. Hexagonal ${\rm YMnO_3}$ is an example of a multiferroic material where the onset of ferroelectricity is completely unrelated to the onset of magnetism, and probably arises from geometrical effects \cite{VanAken}. Orthorhombic ${\rm TbMnO_3}$ is an example of a multiferroic material where ferroelectricity arises from magnetic spiral order \cite{Kimura,KenzelmannTbMnO3,Mostovoy}. Ferroelectricity from magnetic order is related to competing magnetic interactions, whose competition at low temperatures is reduced through small lattice distortions that result in switchable electric polarization.\par

Magnetically induced ferroelectricity has been observed for structurally very different materials, most notably in rare-earth (R) manganites ${\rm RMn_2O_5}$ \cite{Hur,HarrisAharony}, the kagome staircase magnet ${\rm Ni_3V_2O_8}$ \cite{Lawes}, and the triangular lattice antiferromagnet ${\rm RbFe(MoO_4)_2}$ \cite{KenzelmannRbFeMoO}. This suggests that the mechanism to obtain ferroelectricity from magnetic order is quite general and should be present in many materials. In all these materials, ferroelectric polarization arises, at least partly, from incommensurate spiral magnetic structures that lead to polar structures. The ME interaction in these materials is believed to be mediated by spin-orbit interactions, and so the ferroelectric polarization is relatively small.\par

Much larger ferroelectric polarizations were predicted for materials
where ferroelectricity arises from {\it collinear} magnetic order
\cite{SergienkoPRL,Picozzi}. In such materials, ME coupling may be
mediated by the symmetric exchange which is larger than spin-orbit
related interactions. An example is orthorhombic (o) ${\rm HoMnO_3}$
where ferroelectricity arises from commensurate, collinear magnetic
order \cite{Lorenz,MunozHoMnO3}. However, the ferroelectric
polarization in o-${\rm HoMnO_3}$ was observed to be much smaller
than predicted \cite{Picozzi}, and arises partly from rare-earth
magnetic order \cite{Lorenz}. Here, we present the case of
o-${\rm TmMnO_3}$ for which we observed a ferroelectric polarization
that arises from collinear ${\rm Mn^{3+}}$ magnetic order, and that
is at least 15 times larger than observed for o-${\rm HoMnO_3}$. We
provide evidence for spin-lattice coupling effects that are larger
than in other magnetically-induced ferroelectrics.\par

\begin{figure}
\begin{center}
\includegraphics[height=15cm]{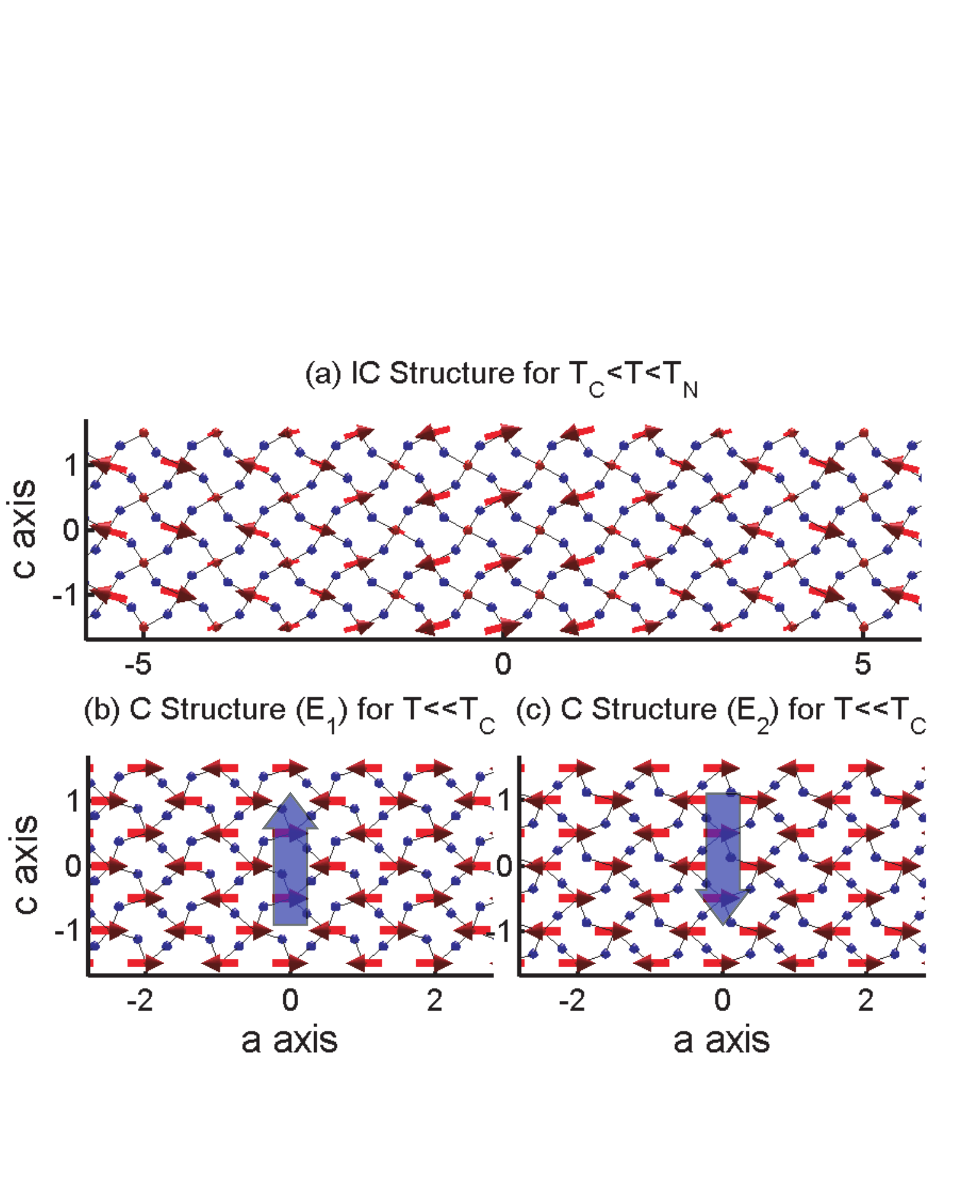}
  \caption{Chemical structure of ${\rm TmMnO_3}$, showing
  ${\rm Mn^{3+}}$ in red and ${\rm O^{2-}}$ in blue.
  (a) Incommensurate amplitude-modulated ${\rm Mn^{3+}}$ spin
  order in the paraelectric phase for $32\;\mathrm{K}<T<40\;\mathrm{K}$.
  (b-c) Commensurate ${\rm Mn^{3+}}$ spin order of $E_1$ and
  $E_2$ type, respectively, in the ferroelectric phase for
  $T \ll 32\;\mathrm{K}$. The large arrows show the direction
  of the spontaneous polarization along the $c$ axis that
  arises from a movement of the ${\rm Mn^{3+}}$ and ${\rm O^{2-}}$
  positions (shown here schematically) to adjust the Mn-O-Mn
  angle for parallel and antiparallel nearest-neighbor alignment,
  thereby lowering symmetry through the creation of
  a polar axis. (a-c) The moments in the neighboring planes
  are oriented in the opposite direction.}
  \vspace{-0.5cm}
  \label{Fig1Structures}
\end{center}
\end{figure}

${\rm TmMnO_3}$ crystallizes in the space group Pnma and has
room-temperature lattice parameters $a=5.809$\,\AA,
$b=7.318\,$\AA$\;$ and $c=5.228\,$\AA. A projection of the crystal
structure onto the $ac$ plane is shown in Fig.~\ref{Fig1Structures}.
The unit cell contains four ${\rm Mn^{3+}}$ ions, located at
$\mathbf{r}_1=(0, 0, 0.5)$, $\mathbf{r}_2=(0.5, 0.5, 0)$,
$\mathbf{r}_3=(0.5, 0, 0)$, and $\mathbf{r}_4=( 0, 0.5, 0.5)$. The large
rotation of the oxygen octahedra around the ${\rm Mn^{3+}}$ ions is
expected to result in appreciable antiferromagnetic superexchange
interactions along the $a$ axis through pairs of oxygen anions
\cite{KimuraPRBC} that compete with the ferromagnetic interactions
in the $ac$ plane.\par

\begin{figure}
\begin{center}
 \includegraphics[height=15cm]{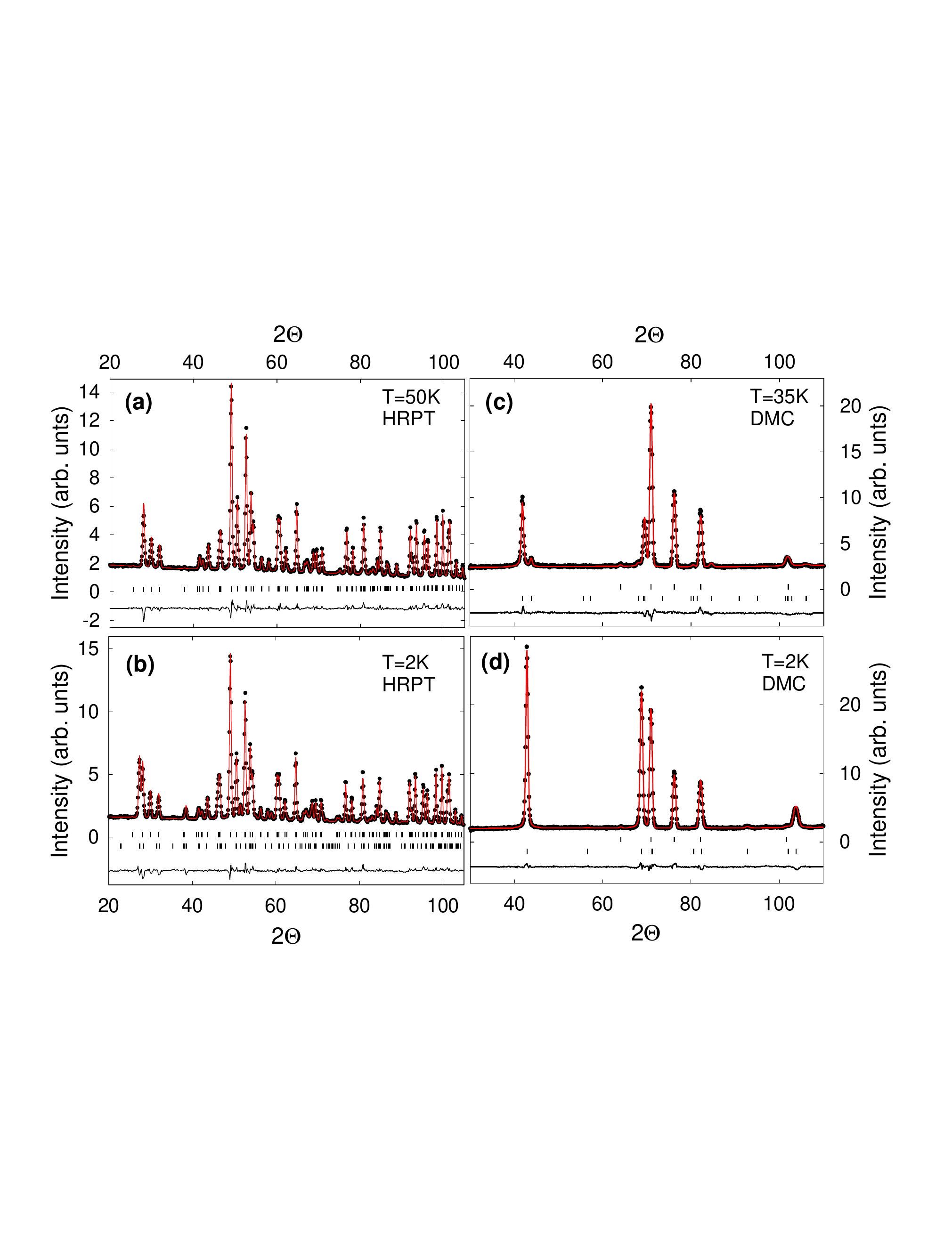}
  \caption{Part of the neutron scattering patterns
  measured using HRPT, as a function of scattering angle
  $2\Theta$ at (a) $T=50\;\mathrm{K}$ showing only nuclear
  scattering, and (b) $T=2\;\mathrm{K}$ showing additionally
  magnetic scattering. (c-d) Bragg peak powder patterns
  measured using DMC at $T=35\;\mathrm{K}$ and $T=2\;\mathrm{K}$.
  (a-d) The vertical bars indicate magnetic and nuclear
  (upper row) Bragg peaks. The bottom solid line indicates the
  difference between the experiment and the model.}
  \vspace{-0.5cm}
  \label{Fig2NeutronPatterns}
\end{center}
\end{figure}

\begin{figure}
\begin{center}
\includegraphics[height=15cm]{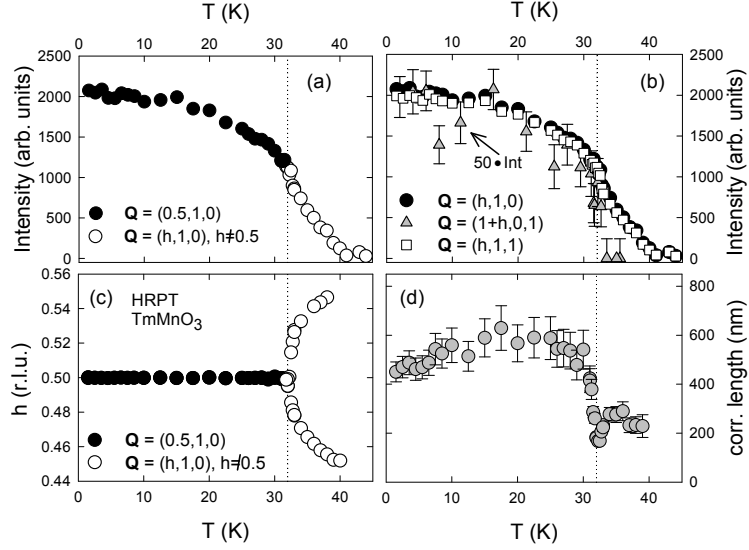}
  \caption{(a) Temperature dependence of the magnetic Bragg
  peak intensity at ${\bf Q}=(0.5,1,0)$ in the commensurate phase,
  or the added intensities at the ${\bf Q}=(q,1,0)$ and
  ${\bf Q}=(1-q,1,0)$ Bragg peak positions for $0.45 < q \leq 0.5$.
  (b) Comparison of the temperature dependence of different magnetic
  Bragg peaks, showing that they have the same temperature
  dependence in the commensurate phase. The ${\bf Q}=(1.5,0,1)$
  Bragg peak is only present in the commensurate phase, and is
  evidence of the ordering of ${\rm Tm^{3+}}$ magnetic moments.
  (c) Temperature dependence of the a-component, $h$, of the
  magnetic Bragg peak ${\bf Q}=(h,1,0)$, where $h=q$ or $h=1-q$.
  (d) Temperature dependence of the magnetic correlation length
  as deduced from the width of the magnetic Bragg peaks.}
  \vspace{-0.5cm}
  \label{Fig3IntQWidth}
\end{center}
\end{figure}

Our neutron diffraction data, shown in Fig.~\ref{Fig2NeutronPatterns}, feature new Bragg peaks below $T^{\rm Mn}_N=42\;\mathrm{K}$ and demonstrate that ${\rm TmMnO_3}$ adopts magnetic order below $T^{\rm Mn}_N$. The ordering wave-vector is ${\bf Q}=(q,0,0)$ where $q$ is the modulation wave-number along the $a$ axis. The temperature dependence of the magnetic neutron Bragg peaks indicates a second-order transition at $T^{\rm Mn}_N$, as shown in Fig.~\ref{Fig3IntQWidth}, and an anomaly at $T_C = 32\;\mathrm{K}$ indicates a further transition. These two
transitions coincide with peaks in the temperature dependence of the
specific heat \cite{Tachibana}. The temperature dependence of the
magnetic peaks close to ${\bf Q}=(0.5,1,0)$
(Fig.~\ref{Fig3IntQWidth}c) shows that the magnetic structure is
incommensurate for $T_C < T < T^{\rm Mn}_N$ and commensurate for
$T<T_C$. In the incommensurate phase, the ordering wave-vector is
${\bf Q}=(q,0,0)$ with $0.45 < q \leq 0.5$.\par

The incommensurate magnetic order is described by one single order parameter, described in more detail in Methods, at $T=35\;\mathrm{K}$ with an amplitude only on the ${\rm Mn^{3+}}$ ions given by ${\bf m}^1_{\rm IC}= \left[ 2.98(2), 0.0(5), \exp(i \phi)\,0.95(3) \right] \mu_B$,
where $\phi$ is the relative phase between the $a$ and $c$- components.
Although we cannot experimentally determine $\phi$, it can be shown
that because of the inversion center of the paramagnetic phase,
$\exp(i \phi) = \pm 1$ \cite{Harris}. No magnetic order was detected on the ${\rm Tm^{3+}}$ ions in the incommensurate phase. Thus the spins are amplitude modulated with moments collinear at an angle to the $a$ axis,
as shown in Fig.~\ref{Fig1Structures}a. This is slightly different from the incommensurate order in ${\rm HoMnO_3}$ that is collinear \cite{MunozHoMnO3}.\par

The commensurate structure at $T=2\;\mathrm{K}$ is described by
two-dimensional order parameter as specified in Methods.
The magnetic order is a E-type magnetic structure shown in
Fig.~\ref{Fig1Structures}b-c, with $3.75(3)\mu_B$ magnetic moment
ordered on the ${\rm Mn^{3+}}$ sites along the $a$ axis. The E-type
magnetic structure can have two independent basis vector for the
moments along the $a$-axis: $E_1=(1,1,-1,-1)$ and $E_2=(1,-1,1,-1)$
in the order of the ${\rm Mn^{3+}}$ ion as defined above - identical
to the low-temperature ${\rm Mn^{3+}}$ order in ${\rm HoMnO_3}$
\cite{MunozHoMnO3}. In addition, we found that ${\rm Tm^{3+}}$ has
an ordered moment of $1.22(5)\mu_B$ pointing along the $c$ axis at
$2\;\mathrm{K}$. Because the ${\rm Tm^{3+}}$ moments are allowed
only along the $b$-axis if they were magnetically polarized by the
${\rm Mn^{3+}}$ order, this implies that the ${\rm Tm^{3+}}$ undergo
independent spontaneous magnetic order, as indicated by a peak in
the specific heat at around $T^{\rm Tm}_N=4\;\mathrm{K}$
\cite{Tachibana}.\par

\begin{figure}
\begin{center}
\includegraphics[height=15cm]{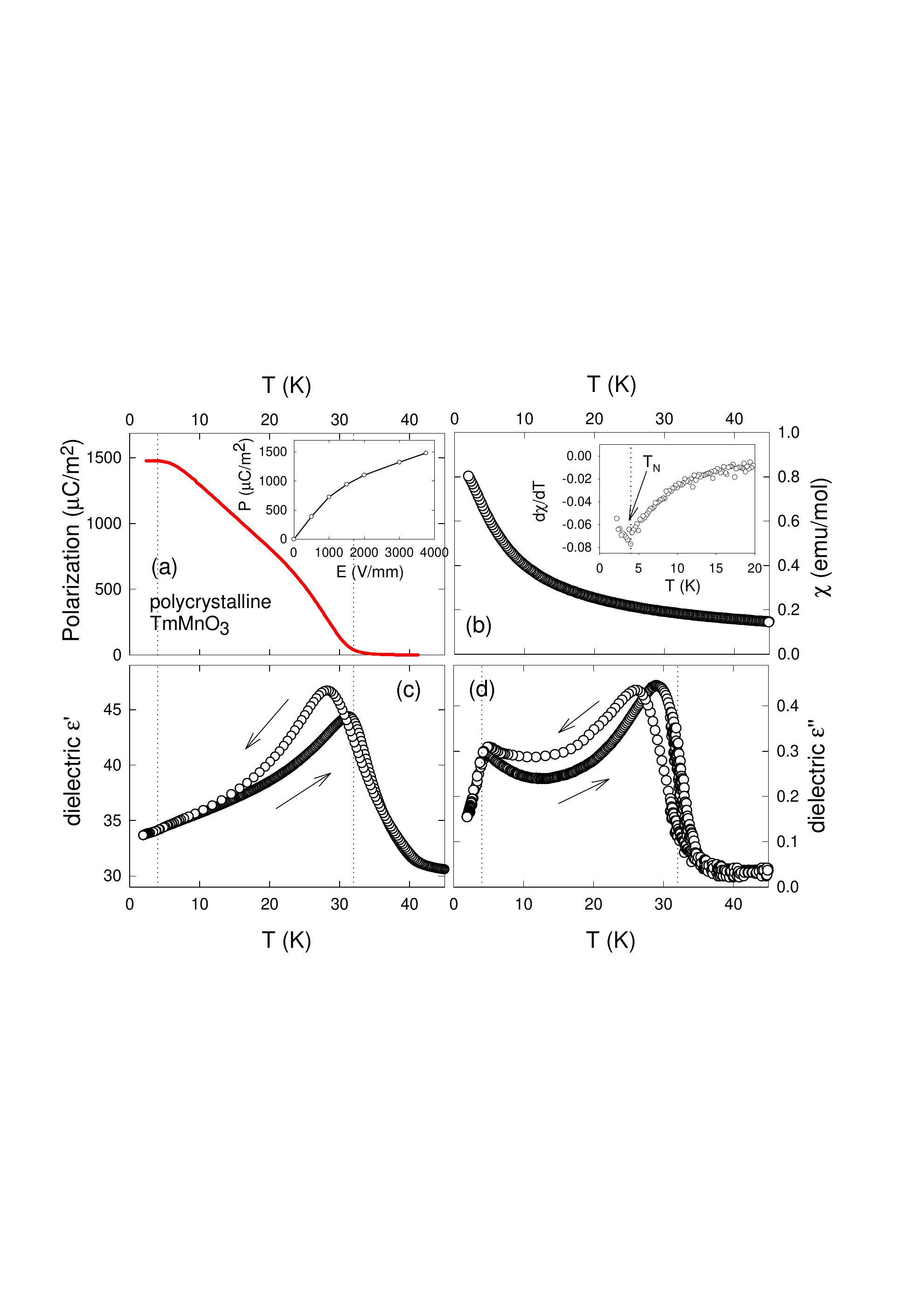}
  \caption{(a) Electric polarization of a pressed powder sample
  of o-${\rm TmMnO_3}$ as a function of temperature, determined
  using pyroelectric measurements after cooling an electrically
  poled sample. Inset: Electric polarization at $T=2\;\mathrm{K}$
  as a function of poling electric field with which the sample
  was cooled. (b) Magnetic susceptibility as a function of
  temperature measured on cooling. Inset: Temperature derivative
  of the magnetic susceptibility indicating the onset of spontaneous
  ${\rm Tm^{3+}}$ magnetic order at $T^{\rm Tm}_N=4\;\mathrm{K}$.
  (c) Real and (d) imaginary part of the dielectric
  susceptibility as a function of temperature,
  measured at a frequency of $f=100\;\mathrm{kHz}$.}
  \vspace{-0.5cm}
  \label{Fig4DielConst}
\end{center}
\end{figure}

Fig.~\ref{Fig4DielConst}a shows that ${\rm TmMnO_3}$ has a
macroscopic response to the onset of magnetic long-range order and
develops spontaneous electric polarization $P$ below
$32\;\mathrm{K}$, demonstrating that o-${\rm TmMnO_3}$ has a
multiferroic ground state. The observed value of $P$ for a powder
sample, $P = 1500\;\mathrm{\mu C/m^2}$, is more than 15 times larger
than that of o-${\rm HoMnO_3}$ \cite{Lorenz}. The value of $P$ for a
powder sample is half the intrinsic value for a single crystal,
namely $P_0=0.3\;\mathrm{\mu C/cm^2}$. Since we have not observed
the saturation of $P(E)$, as shown in the inset of
Fig.~\ref{Fig4DielConst}a, $P_0$ may be substantially higher and our
observation is a lower limit of the intrinsic polarization. The
reported electric polarization in o-${\rm HoMnO_3}$ was much
smaller, so our results suggest that sample quality or the details of the crystal structure are decisive for the size of the electric polarization in the orthorhombic rare earth manganites. The experimentally observed
polarization (which is merely a lower limit for the intrinsic electric polarization) is the highest observed value for magnetically induced
ferroelectricity to date, and is of the same order as the values
$P_0=0.5-12\;\mathrm{\mu C/cm^2}$ \cite{SergienkoPRL} and
$6\;\mathrm{\mu C/cm^2}$ \cite{Picozzi} predicted (but not observed) for ${\rm HoMnO_3}$. This provides strong experimental evidence that the
theoretically predicted mechanism of symmetric exchange,
although not universal to all o-${\rm RMnO_3}$ systems, does
apply in the case of ${\rm TmMnO_3}$ and can give rise to
magnetically-induced ferroelectricity that is large enough for
applications.\par

From the magnetic structures shown in Fig.~\ref{Fig1Structures} we
propose a likely scenario for the magnetic exchange interactions in
${\rm TmMnO_3}$. These structures suggest that the interactions
between second neighbors are ferromagnetic along the $c$ axis and
are antiferromagnetic along the $a$ and $b$ axes. In the
commensurate phase (for $T<T_C$) the distortion of the nearest
neighbor bonds is such that the straighter bonds have an interaction
that is less ferromagnetic (or more antiferromagnetic) than the bent
bonds, thus removing the frustration that would occur in the absence
of the distortion. For $T_C<T$, when the bonds are undistorted, the
frustration is removed by the incommensurate structure of
Fig.~\ref{Fig1Structures}a.\par

The magnetic order is never strictly long-range, because magnetic
Bragg peaks were found to be always wider than the
resolution-limited nuclear Bragg peaks. Fig.~\ref{Fig3IntQWidth}d
shows that the magnetic correlation length does never exceed
$600\;\mathrm{nm}$, and most probably arises from ferroelectric
domains. Picozzi {et. al.} \cite{Picozzi} showed that ferroelectric
polarization in ${\rm HoMnO_3}$ is generated mostly through
movements of the ${\rm Mn^{3+}}$ and ${\rm O^{2-}}$ positions, so
the magnetic structure E$_1$ and E$_2$ (shown in
Fig.~\ref{Fig1Structures}) favor opposite ferroelectric
polarization, as can be seen from the phenomenological formula $P_0
\propto (E_1^2-E_2^2)$. Thus the magnetic structure E$_1$ and E$_2$
must be separated by a magnetic domain walls, limiting the magnetic
correlation length to the size of the ferroelectric domains. Our
measurements thus suggest that the magnetic domains can be
controlled by electric fields.\par

The temperature dependence of the real part of the dielectric
susceptibility, shown in Fig.~\ref{Fig4DielConst}c provides evidence
for the ferroelectric transition at $T_C=32\;\mathrm{K}$, in
agreement with the pyroelectric measurements. The imaginary part of
the dielectric constant, shown in Fig.~\ref{Fig4DielConst}d, shows a
two-peak feature as a function of temperature, and relatively high
values between $T^{\rm Tm}_N=4\;\mathrm{K}$ and $T_C=32\;\mathrm{K}$
that suggest substantial energy dissipation. The energy dissipation
in this temperature range may result from slow switching behavior
associated with the magnetically polarized ${\rm Tm^{3+}}$ magnetic
moments that are only loosely coupled to the rapidly switching ${\rm
Mn^{3+}}$. Below $T^{\rm Tm}_N=4\;\mathrm{K}$, the ${\rm Tm^{3+}}$
moments are spontaneously ordered and therefore not directly
connected to the electric order, so that dielectric constant shows
no dissipation, as shown in Fig.~\ref{Fig4DielConst}d. This scenario
is also consistent with a flattening off of the electric
polarization stops below $T^{\rm Tm}_N=4\;\mathrm{K}$, suggesting
that the ${\rm Tm^{3+}}$ order competes with the ${\rm Mn^{3+}}$
order and thereby limits the size of the electric polarization.\par

\begin{figure}
\begin{center}
\includegraphics[height=15cm]{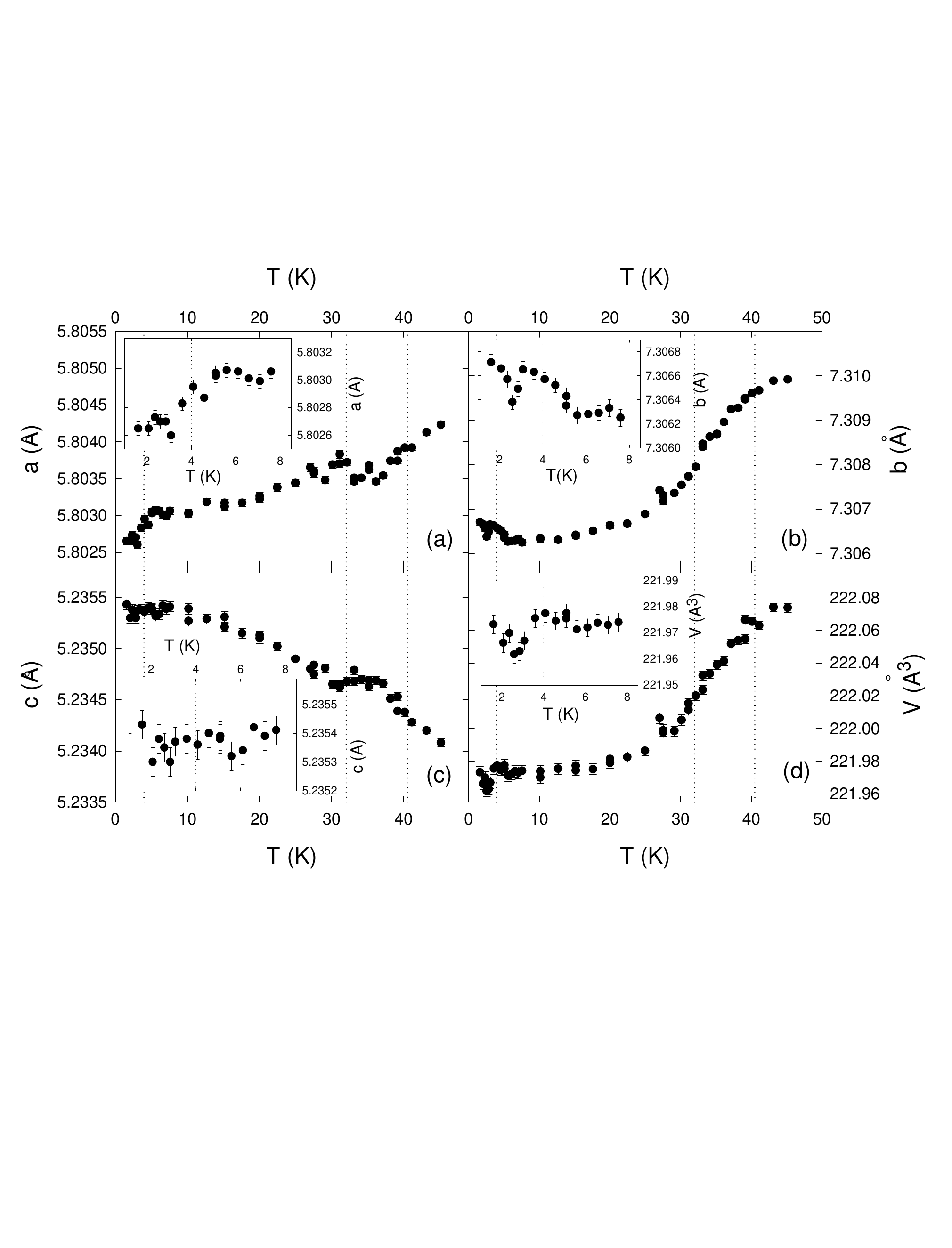}
  \caption{Temperature dependence of the lattice constants for
  $T<50\;\mathrm{K}$. The insets show additional transitions
  below $4\;\mathrm{K}$. The
  vertical dotted lines at $T^{\rm Mn}_N$ and $T_C$ indicate magnetic
  transitions, while the vertical dotted line at
  $T^{\rm Tm}_N=4\;\mathrm{K}$
  indicates the onset of spontaneous ${\rm Tm^{3+}}$ magnetic order.}
  \vspace{-0.5cm}
  \label{Fig5LattConst}
\end{center}
\end{figure}

Independent evidence for strong coupling between the chemical and
magnetic lattice is also seen in the temperature dependence of the
lattice constants, shown in Fig.~\ref{Fig5LattConst}. These
spin-lattice effects are larger than in any other heavy rare-earth
o-${\rm RMnO_3}$, suggesting that the magnetic order has a stronger
effect on the chemical lattice of o-${\rm TmMnO_3}$ than in other
heavy rare-earth manganites. Our results can be understood
phenomenologically as follows. Because the incommensurate magnetic
order is described by only a single one-dimensional order parameter,  there can be no magnetically-induced
ferroelectricity in accordance with our experiment \cite{Harris}. In
the commensurate phase the ME interaction is of the form given in
Ref.~\onlinecite{SergienkoPRL}. However, the fourth order terms in
the magnetic free energy cause either $E_1 \cdot E_2=0$ or
$|E_1|=|E_2|$, depending on the sign of the fourth order spin
anisotropy\cite{Harris08}. Thus the higher order ME interaction in
Ref.~\onlinecite{SergienkoPRL} is generally inoperative and the
polarization is restricted to lie along the $c$ axis with magnitude
$P_c \propto (E_1^2-E_2^2)$, where $E_1E_2=0$ is selected. The
temperature dependence of $P$ is only qualitatively consistent with
this, possibly because the results are somewhat modified by the
sample not being a single crystal.\par

In summary, we have shown that ${\rm TmMnO_3}$ has a
magnetically-induced electric polarization that is substantially
higher than in any other heavy rare-earth manganites with
commensurate magnetic order. We observed anomalies in the
temperature dependence of the lattice constants at the magnetic
phase transitions that are evidence for strong coupling effects
between the chemical and magnetic lattices. Theoretical calculations
have predicted a large spontaneous electric polarization in ${\rm
HoMnO_3}$, at variance with current experimental results
\cite{Picozzi}. Since we have found such a large polarization in
${\rm TmMnO_3}$, it is of great interest to have such calculations
made for this system and hopefully to understand the difference
between ${\rm HoMnO_3}$ and ${\rm TmMnO_3}$.\par

\begin{acknowledgments}
We acknowledge valuable discussions with R.~A. Cowley, N.~A.
Spaldin, and D. Khomskii. This work was supported by the Swiss NSF
(Contract No. PP002-102831). This work is based on experiments
performed at the Swiss spallation neutron source SINQ, Paul Scherrer
Institute.
\end{acknowledgments}

\vspace{1cm}
{\bf Methods}\\

Polycrystalline samples of perovskite ${\rm TmMnO_3}$ were prepared
under high pressure as described in Ref.~\onlinecite{Tachibana}. Neutron powder diffraction measurements were performed on a large
amount ($5.4\;\mathrm{g}$) of ${\rm TmMnO_3}$ sample using the HRPT
and DMC diffractometers at the Paul Scherrer Institute, and incident
neutrons with a wave-length of $1.89\;$\AA$\;$ and $4.5\;$\AA,
respectively. The magnetic structures were determined using the Fullprof Suite \cite{Fullprof}. The size of the magnetic moments have been determined by comparing the strength of magnetic and nuclear intensities. No texture effects were observed during the analysis.\par

The ferroelectric polarization was determined using a $0.4\;\mathrm{mm}$ thin hardened pellet of polycrystalline ${\rm TmMnO_3}$ covered with an area $3.12\cdot10^{-6}\;\mathrm{m^2}$ of silver epoxy. The sample was cooled from $50\;\mathrm{K}$ to $2\;\mathrm{K}$ in poling electric fields of up to $E=3750\;\mathrm{kV/m}$, after which the electric field was reduced to zero and the sample was allowed to discharge for $5$ minutes.
After the discharge at $2\;\mathrm{K}$ the residual current was reduced to $10^{-14}\;\mathrm{A}$, which suggests that trapped charges did not affect the pyroelectric measurement. Then the sample was heated at different constant rates between $0.85$ and $4.86\;\mathrm{K}/{\rm min}$ and the pyroelectric current was measured using a Keithley 6517A electrometer, resulting an nearly identical estimates of the ferroelectric polarization. Pyroelectric measurements at different ramping speeds and a stop-and-go ramp result in a nearly identical temperature dependent electric polarization, showing the thermal excitation of trapped charges does not affect the pyroelectric measurements. These measurements therefore allow the determination of the lower limit of the electric polarization. Real and imaginary part of the dielectric constant were measured using a Agilent E4980A LCR meter, making sure that the Maxwell-Wagner effect does not affect the measurements. The magnetic susceptibility was measured in an external field $H = 100 \;\mathrm{Oe}$ on a small ($5.9\mathrm{mg}$) powder sample using a Quantum Design SQUID magnetometer.\par

The incommensurate magnetic structure belongs to
irreducible representation $\Gamma_{\rm IC}^3$, where the
superscript corresponds to Kovalev's notation \cite{Kovalev}, and is
defined by the following characters: $\chi(I)=1$,
$\chi(2_a)=-\alpha$, $\chi(m_{ab})=-\alpha$ and $\chi(m_{ac})=1$,
with $\alpha=\exp(i\pi q)$. Here $2_a$ is a two-fold screw axis
rotation, while $m_{ab}$ and $m_{ac}$ are $ab$/$ac$-mirror planes
followed by a $(0.5,0,0.5)$ or $(0,0.5,0)$ lattice translation,
respectively. The commensurate structure at $T=2\;\mathrm{K}$ is
described by the two-dimensional irreducible representation $\Gamma_{\rm C}^1$ according to Kovalev's notation and defined by the following
non-zero characters: $\chi(I)=2$ and $\chi(m_{ac})=-2$.\par

\end{document}